# A few misfits can Change the World

Esteve Almirall[1] esteve.almirall@esade.edu,
Steven Willmott[2], steve@timewarp.io and
Ulises Cortés[3] ia@cs.upc.edu

[1] Esade, Universitat Ramon Llull, Av. Pedralbes 60, 08034 Barcelona
[2] TimeWrap Labs, Austin, Texas
[3] UPC, Universitat Politècnica de Catalunya, c/ Jordi Girona 31, 08034, Barcelona

**Abstract:** Rising inequality is a critical concern for societies worldwide, to the extent that emerging high-growth economies such as China have identified *common prosperity* as a central goal. However, the mechanisms by which digital disruptions contribute to inequality and the efficacy of existing remedies such as taxation, must be better understood. This is particularly true for the implications of the complex process of technological adoption that requires extensive social validation beyond weak ties and, how to trigger it in the hyperconnected world of the 21[st] century.

This study aims to shed light on the implications of market evolutionary mechanism through the lenses of technological adoption as a social process. Our findings underscore the pivotal importance of connectivity in this process while also revealing the limited effectiveness of taxation as a counterbalance for inequality. Our research reveals that widespread cultural change is not a prerequisite for technological disruption. The injection of a small cohort of entrepreneurs – a few misfits – can expedite technology adoption even in conservative, moderately connected societies and, change the world.

**Keyboards:** Inequality, Technological Disruption, Economic Growth, Agent-Based Model

## Introduction

The increasing disparities in economic wealth and opportunities (Piketty 2019; Acemoglu and Robinson, 2012; Milanovic 2016; Atkinson 2015; Stiglitz 2012 ), represent a pressing challenge for contemporary societies. This challenge is accentuated by the ascendance of digital multinationals, whose extensive influence transcends traditional boundaries fostering globalization. Concurrently, the ongoing digital transformation threatens to potentially exacerbate these disparities. The question if inequality is an avoidable outcome of digital disruptions and if so, potential remedies, becomes increasingly pertinent. As new disruptions such as generative A.I. loom on the horizon, understanding the underlying mechanisms assumes an even greater significance.

This manuscript aims to explore the processes through which digital disruptions engender social inequality through a process of technological adoption. In particular, technological adoption that is more realistic than typical models generally used and requires extensive social validation for technology adoption rather than just simple contact (Centola 2007).

Inequality's roots are undoubtedly complex, resulting from a myriad of factors, including power conflicts among varying societal factions, inherent benefits such as oil resources, economies of agglomeration, and several others. Yet, there has been a long-standing pursuit to identify internal factors and broad principles that could elucidate the origins of inequality. The initial effort to do so was made by Pareto in 1897, who suggested that inequality follows a ubiquitous power law applicable across all time periods and countries. This hypothesis, however, faced contention (Shirras 1935), and subsequently, Mandelbrot proposed a modified Pareto law (1960), which was primarily applicable to high-income brackets. Over time, numerous other distributions have been proposed (Kakwani 1980).

The foundational reasoning for these proposals diverges into two distinct schools of thought. One school promotes socio-economic justifications (Levy 1987), while the other interprets it as a stochastic or random process. For instance, Gribrat (1931; Montroll and Shlesinger 1983) asserted that inequality is a product of a multiplicative random process. In contrast, Kalecki (1945) postulated that the variance expands over time, and Levy and Solomon proposed a lower income cut-off point, thereby stabilizing the distribution into a power law (1996).

However, some posit that the core issue lies in the dynamics of wealth creation and market structure, tailored for an economic landscape starkly different from our current digital era (Piketty 2019; Acemoglu and Robinson 2012; Milanovic 2016),

In the past, markets typically followed a Gaussian distribution, where the majority of businesses experienced average success, with a few outliers at the extremes of success and failure. Traditional enterprises, such as bakeries, restaurants, and conventional grocery stores, typically exhibit this type of distribution. The reason lies in inherent constraints – the best bakery in the world can only cater to nearby residents, with expansion possible only by launching another bakery, thus resulting in linear growth. These limitations exist not only in the business model but also in the operational model. Physical products usually demonstrate decreasing returns to scale, and while the threshold for these returns has expanded with the advent of mega-factories like Tesla's, they remain present.

On the contrary, digital markets are free from these traditional boundaries. If one can develop the world's leading search engine, it becomes instantly accessible to everyone, everywhere (Shapiro & Varian 1999). Moreover, digital products do not display decreasing returns; rather, they often exhibit increasing returns (Arthur 1996), facilitated, in part, by the persistent relevance of Moore's law (Moore 1965).

These conditions have cultivated a distinctive culture among digital startups, characterized by high growth multipliers substantially exceeding those of traditional firms. Where S&P 500 companies demonstrate an average growth rate of approximately 10%-15%, startups are expected to achieve minimum growth rates ranging from 40% to nearly 200%, resulting in earning multipliers of 20, 50, or even 150. This dynamic has engendered a shift in market portrayal from Gaussian to power law distributions, alongside a culture of rapid growth, pivoting, and the 'J-curve' phenomenon (Kaplan and Schoar, A. 2005; Phalippou and Gottschalg, 2009; Robinson and Sensoy 2016)

Inequality is multifactorial and often emerges from power dynamics, differential access to scarce resources, or even, sheer serendipity. This paper aims to examine inequality from an evolutionary standpoint as a result of technological disruptions in the context of increasing globalization. We aim to validate the following postulates:

> H1.- Disparities in growth trajectories between digital-centric and traditional enterprises are huge drivers of inequality. Evolutionary amplification heavily compounds this difference.

> H2.- Progressive taxation is not an effective counterbalance to the inequality caused by such digital disruptions unless taken to unrealistic extremes.

> H3.- High Levels of Network Connectivity Are Essential for Technological Disruptions.

> H4.- A minimal entrepreneurial seeding can catalyze extensive technological adoption.

In our exploration we initially concentrate on exploring how markets, functioning as evolutionary mechanisms, amplify existing growth disparities and assess the capability of progressive taxation as a counterbalance against inequality. Subsequently, we shift our focus to technological adoption, considering it as a socially-mediated process where inequality emerges not as an exogenous factor but as a property of the system, contingent upon its level of connectivity.

As our inquiry progresses, we find that full adoption of a technological disruption largely hinges on societal validation (Centola 2007), thus being dependent on a community's proclivity for technology adoption and risk-taking. Remarkably, our analysis shows that inducing wide-scale adoption does not require wholesale cultural shifts; rather, the introduction of a small cohort of risk-prone entrepreneurs - 'a few misfits' - can suffice to catalyze broad technological uptake.

By distilling these insights, this study aims to enrich the current discourse surrounding digital adoption, technological transformation, societal inequality, and agent-based modeling of technology diffusion. Additionally, we hope that our findings will provide nuanced perspectives that can inform more effective policymaking, therefore, better policies.

## Markets as an evolutionary mechanism

To address our first hypothesis, we will first use a highly stylized simulation model to explore the effects of diverse growth rates in an evolutionary context. By isolating this specific aspect of the problsem, our objective is to discern the mechanisms through which, and the extent to which, diversity in growth rates becomes amplified and its impact on inequality.

In this initial model, our primary objective is to model the role played by the evolutionary dynamics of markets. As such, allocating digital and traditional companies is predetermined and exogenous to the model's design. The simulation encompasses a series of agents denoted as

A={$a_1$, …, $a_n$}. These agents epitomize individual firms, each endowed with a specific valuation that undergoes evolution as time progresses. Notably, these firms are classified into one of two growth trajectories: traditional or digital companies.

A distinguishing factor between these two company types is their respective growth rates. Traditional companies manifest a growth rate characterized by a Gaussian distribution with a mean of 0.15 (or 15%) and a standard deviation ($\sigma$) of 0.1. This rate is analogous to the enduring growth rate of the SP500. In contrast, digital companies are modeled to mirror the growth patterns of digital startups, commencing with a robust growth rate of 150%, which then depreciates over a decade to align with the growth rates exhibited by traditional companies.

It is imperative to note that once companies have their valuation sink below a single unit, they are officially deemed bankrupt, paving the way for their replacement by a nascent entity.

### Agent Behavior

Agents are initialized with a minimal initial valuation. Their classification – whether traditional or digital – is determined through a random selection, influenced by a predetermined probability scale ranging in the interval [0 … 1]. Here, a probability of 0 implies all agents are traditional, whereas a probability of 1 denotes all agents as digital.

During each iteration of the simulation:

1. Agents are activated in a stochastic manner, and their valuation either appreciates or depreciates based on their designated type.

2. Traditional agents experience fluctuations in their valuation derived from profits or losses, which are determined by a percentage drawn from a normal distribution with a mean of 0.15 and a standard deviation ($\sigma$) spanning from 0.1 to 1.

3. For digital agents, their growth coefficient is procured randomly from a normal distribution that has both a mean and $\sigma$ set at one.

4. Digital companies experience a consistent decline in their growth rate which is depreciated at every simulation iteration. Their eventual growth rate is discerned by deducting twice the growth rate of traditional companies from the digital companies' rate.

Should an agent's valuation be entirely depleted, it is considered *defunct* and subsequently substituted by a *fresh* agent. This new agent type (traditional or digital) is allocated through through random selection, guided by the specific probability set for that simulation run. Consequently, the total number of agents remains unaltered.

Agents must to pay taxes on their accrued profits on the financial front. The tax model is versatile: it can be depicted as a uniform rate or adopt a progressive structure with varying rates set per quantile. These quantiles are determined based on the valuation distribution of the entire agent population during each simulation step.

## Computational experiments

Our computational experiments involved simulating 10,000 agents across ten periods, conducting a thousand iterations for each period. The results we discuss are the averaged outcomes of these iterations.

All simulations were coded in Julia 1.9, harnessing the functionality of the agents.jl library (Datseris, Vahdati, and DuBois, 2021).

For a detailed breakdown of the parameters governing our simulations, readers are directed to Table 1.

| PARAMETER | VALUE |
|---|---|
| Number of agents | 10,000 |
| Number of Periods | 10 |
| Number of experiments | 1,000 |
| Taxes | 15% flat |
|  | 10% 15% 25% 30% per quantile |
|  | 15% 25% 50% 70% per quantile |
| Profits/loses per period | a random number from a normal distribution of $\mu = 0.15$ and $\sigma = \{0.1..1\}$ for traditional companies and $\mu = 1.5$ and $\sigma = 1$ for digital companies |
| Depreciation for digital companies | 10 periods at a rate of 1/10 per period Final growth rate = growth rate of digital companies – two times the growth rate of traditional companies |
| Base growth rate of digital companies after depreciation | Two times the one of normal companies |
| Initial valuation | Ten units |
| Percentage of digital companies | 0..1 |

*Table 1. Parameters of the simulation*

## Results

In our initial experiments, we see that a low standard deviation (0.1) in the percentage of growth (profit and loss) across traditional organizations results in a distribution of gains that resembles a Gaussian distribution, as might be anticipated. As expected, as the standard deviation ($\sigma$) increases, the distribution shifts. It may come as a bit of a surprise, that even if just traditional companies, it becomes one more reminiscent of a power-law distribution, increasing inequality.

Therefore, we already see that greater diversity in gains and the selection market mechanism drive this transformation, increasing inequality steadily.

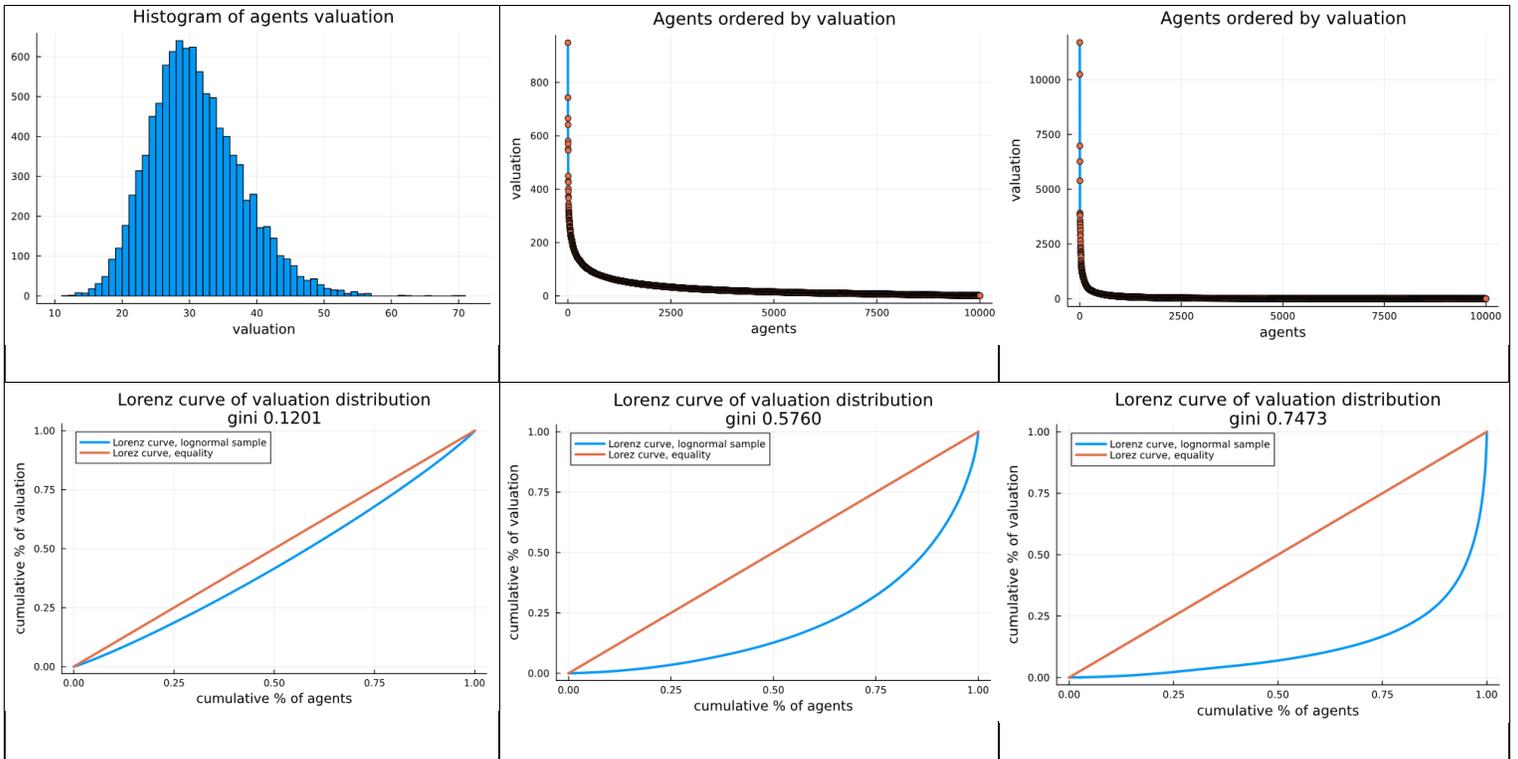

*Figure 1. Agents' distribution for a σ of growth rate for gains/losses of 0.1, 05 and 1*

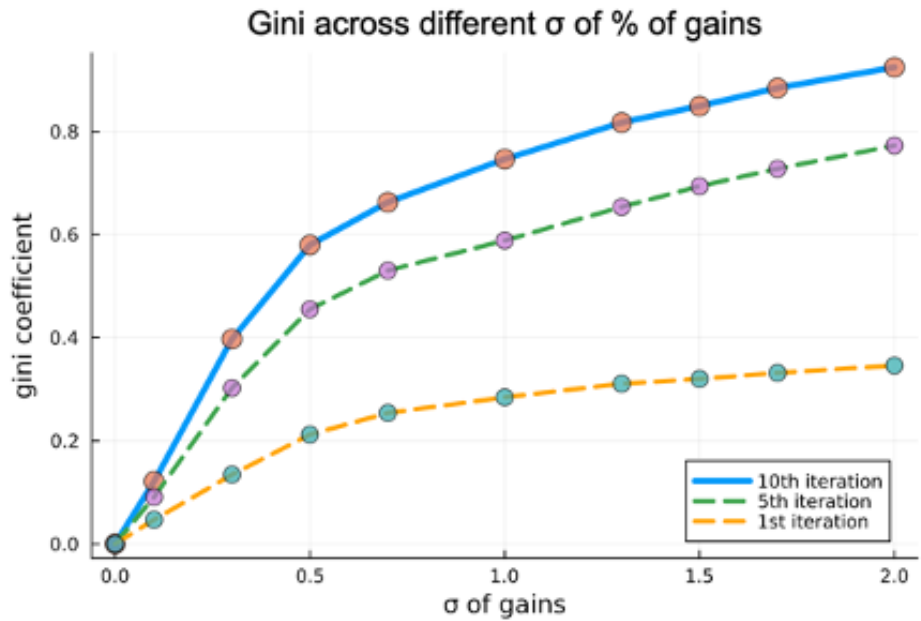

*Figure 2. Increase of inequality as diversity in the distribution of gains increases*

The amplifying effect of markets' evolutionary nature on existing disparities is vividly illustrated in Figure 2. Here, it becomes evident that within a limited number of iterations, a modest Gini coefficient escalates to significant levels.

For our second round of experiments, we explore this mix of traditional and digital companies, ranging from a very small percentage of digital companies (1%) to 30% and 100%.

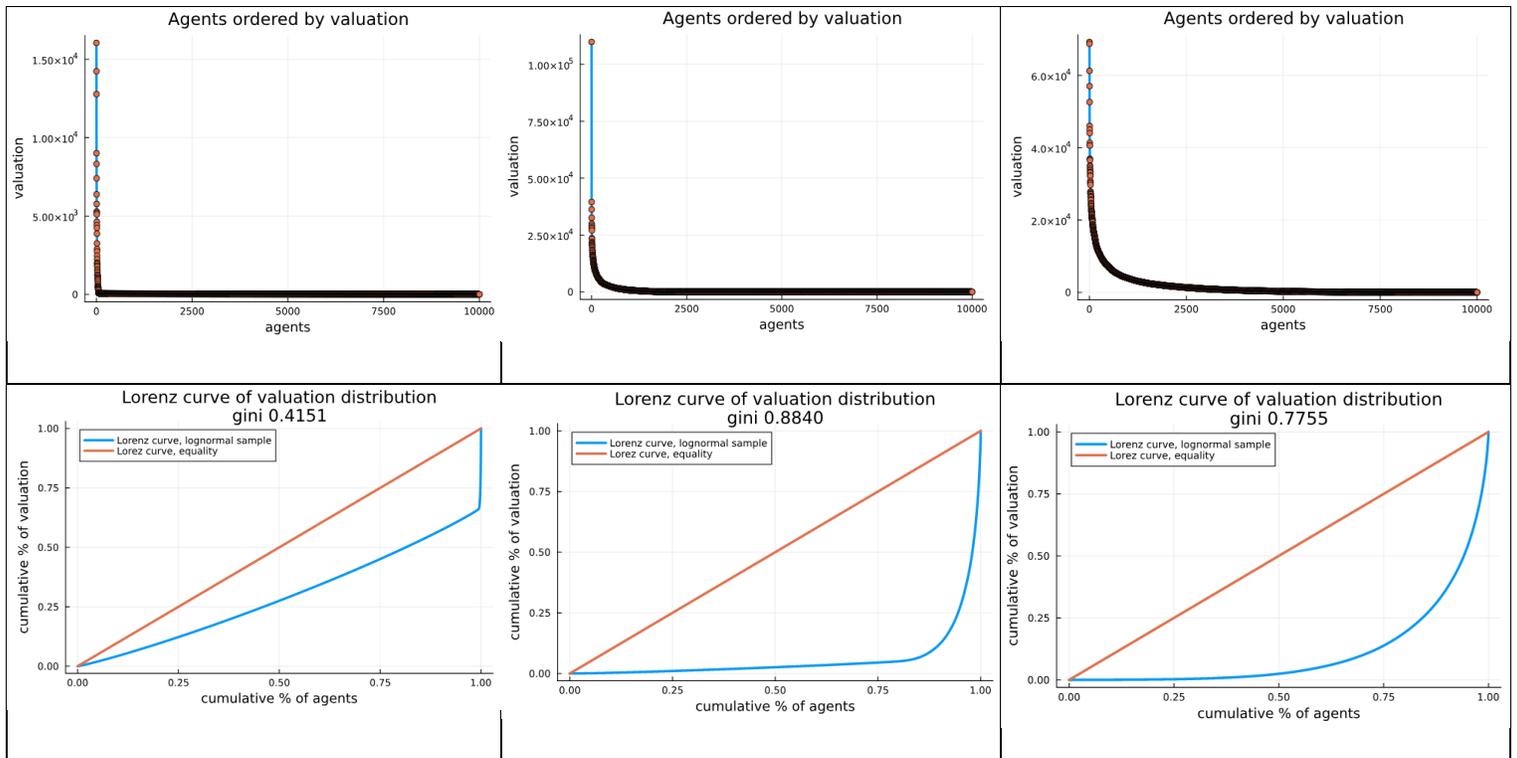

*Figure 3. Inequality in a mix of digital and traditional companies (0.01, 0.3, 1)*

In our experiments, we have thus far considered a static framework encompassing two distinct categories of companies: traditional businesses, with a growth rate randomly drawn from a normal distribution $N(0.15, 0.1)$, and digital businesses, where the growth rate is randomly sampled from $N(1.5, 1)$, corresponding to the average growth of SP500 companies and the common expected multiplicator for digital startups in accelerators. However, it is imperative to recognize that this static portrayal will undergo transformations as digital elements progressively infiltrate traditional enterprises. To capture this evolution, we adjusted the standard deviation σ for traditional companies by randomly sampling it from the positive segment of a normal distribution ranging from *N(0, 0.1)* to *N(0, 1)*. Consequently, each company now exhibits a distinct σ value for its growth rate. In contrast, the mean μ of these distributions (from which the standard deviation of the growth rate is derived) progresses in tandem with their digital metamorphosis.

During the initial phase, when the standard deviation of growth is extracted from *N(0, 0.1)*, our analysis reveals that inequality reaches its zenith around a composition of 30% digital companies and subsequently declines, albeit gradually.

Nevertheless, as the digital transformation progresses, so too does the inequality. Our observations indicate that the inequality's contribution from traditional companies, which increasingly adopt digital characteristics, rises in proportion to the standard distribution. An unmistakable inverted U-curve emerges in this process. Specifically, for cases where σ is randomly chosen from *N(0, 0.5)* or less, digital firms are the primary contributors to this inequality.

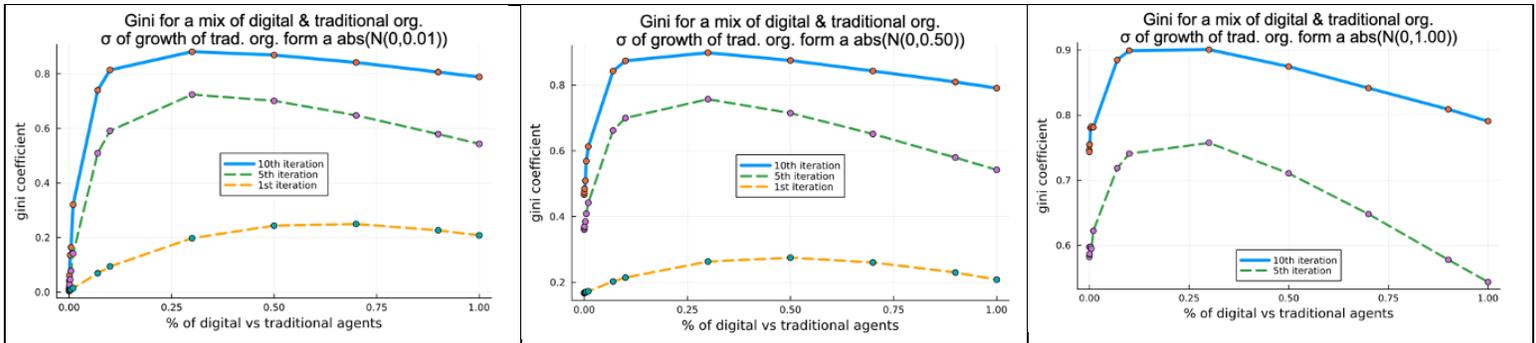

*Figure 4. Gini for a mix of digital and traditional organizations*

Once again, the data shows how the evolutionary characteristics of markets exacerbate modest levels of inequality, reaching heightened proportions and thereby refuting our initial first hypothesis.

## Can progressive taxation solve inequality?

Progressive taxation is the conventional tool that societies utilize to mitigate inequality. In our study, we sought to evaluate the effectiveness of this tool concerning two sources of inequality: one stemming from an increased standard deviation in the percentage of gains from traditional companies and the other from the dichotomy between two different types of organizations in terms of growth, namely traditional and digital.

To gauge the impact of progressive taxation, we established two taxation scales, moderate and high, and applied them to the inter-quantile space of the gains experienced by agents in the preceding period (for the first period, we set quantiles at 10, 20, and 30). We used standard quantiles of 0.25, 0.5, and 0.75 in the process.

| quantiles | 0.25 | 0.5 | 0.75 | |
|---|---|---|---|---|
| Taxation – moderate | 10% | 15% | 25% | 30% |
| Taxation – high | 15% | 25% | 50% | 70% |
| Taxation – extra high | 10% | 15% | 30% | 90% |

*Table 2. Taxation rates moderate and high*

In the context of traditional companies, our observations reveal that progressive taxation effectively curtails inequality, which rises only up to a standard deviation of gains of σ = 0.5 and remains stable beyond that point. The only discernible difference between the two taxation scales is the resultant level of inequality, which diminishes as taxation rates escalate.

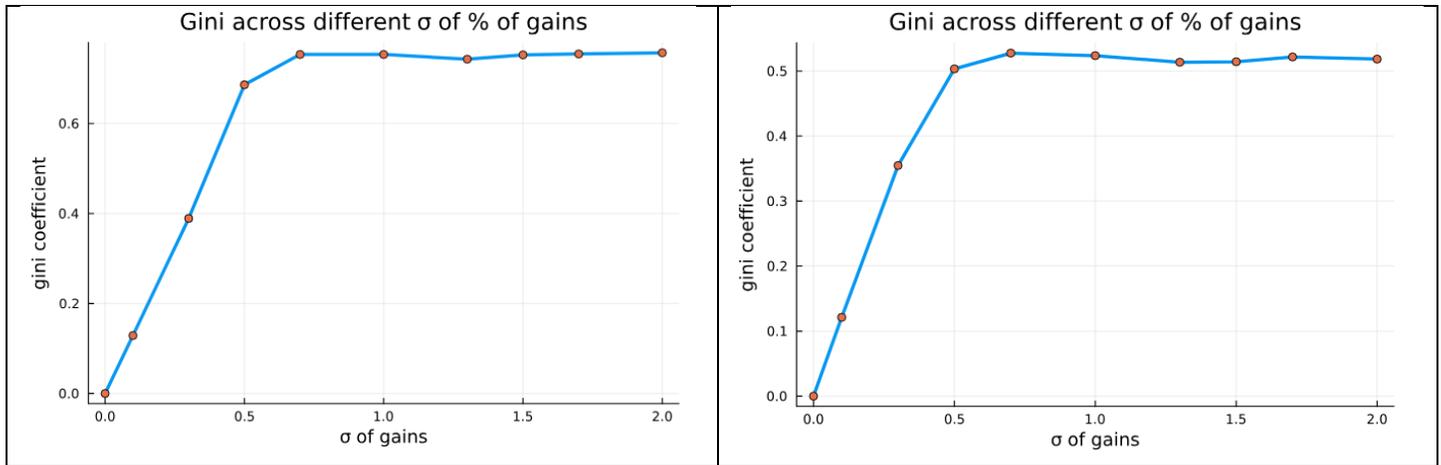

*Figure 5. Gini across different σ for taxes 10%,15%,25%, 30% and 15%, 25%, 50%, 70%*

This finding is hardly surprising, as progressive taxation has evolved as an instrument to combat inequality in societies where traditional companies, constrained by decreasing returns on scale, have been the norm.

In contrast, when examining a blend of traditional and digital organizations with different types of growth, we can identify notable distinctions.

With a moderate taxation rate, the impact on inequality is somewhat subdued. While it manages to decelerate the advancement of inequality when the percentage of digital companies is minuscule, and slightly diminishes the overall inequality in all scenarios, the curve's shape and progression stay largely consistent.

However, when the taxation rate is high, the effects become more pronounced. High taxation succeeds in tempering the progression of inequality, flattening it to levels that are analogous to those seen in traditional companies. Yet, when taxation in the top quartile becomes extremely and unrealistically high in an attempt to prevent inequality, we observe that the evolutionary nature of markets triumphs, letting only the top survive and abruptly and unexpectedly increasing the level of inequality as the number of digital high-risk, high-growth agents increases.

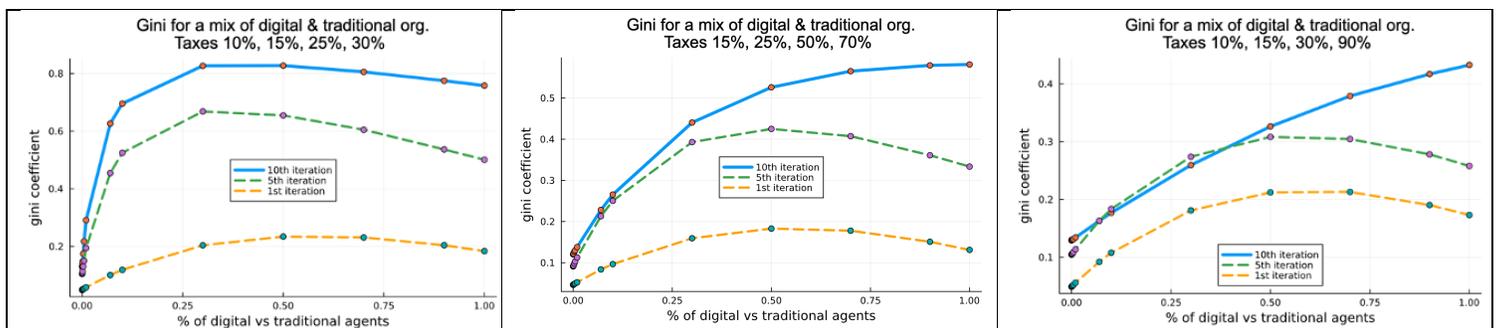

*Figure 6. Gini for a mix of traditional and digital companies for taxations of 10%,15%,25%, 30%, 15%, 25%, 50%, 70% and 10%, 15%, 30%, 90%.*

We can further extend our analysis by examining the impact of taxation, particularly focusing on the escalated standard deviations in the distributions of growth across individual agents.

These simulations reveal the interplay between the escalation in inequality, driven by increasing values of the standard deviation of growth, and the dual influences of the evolutionary dynamics of markets and progressive taxation. While these effects resonate with our previous observations, there is an added layer of complexity due to the amplification of inequality resulting from the elevated standard deviation of growth.

In both scenarios, whether assessing the effects of digital transformation or progressive taxation, the emergence of the inverted U-shaped curve remains consistent, punctuated by a tipping point of inequality. This tipping point represents a critical juncture in the model, highlighting a transitional phase where various factors may weigh differently on the resulting inequality dynamics. It emphasizes the subtle balance and interdependence of factors in the complex system of digital transformation, potentially offering pivotal insights for policy interventions and strategic considerations.

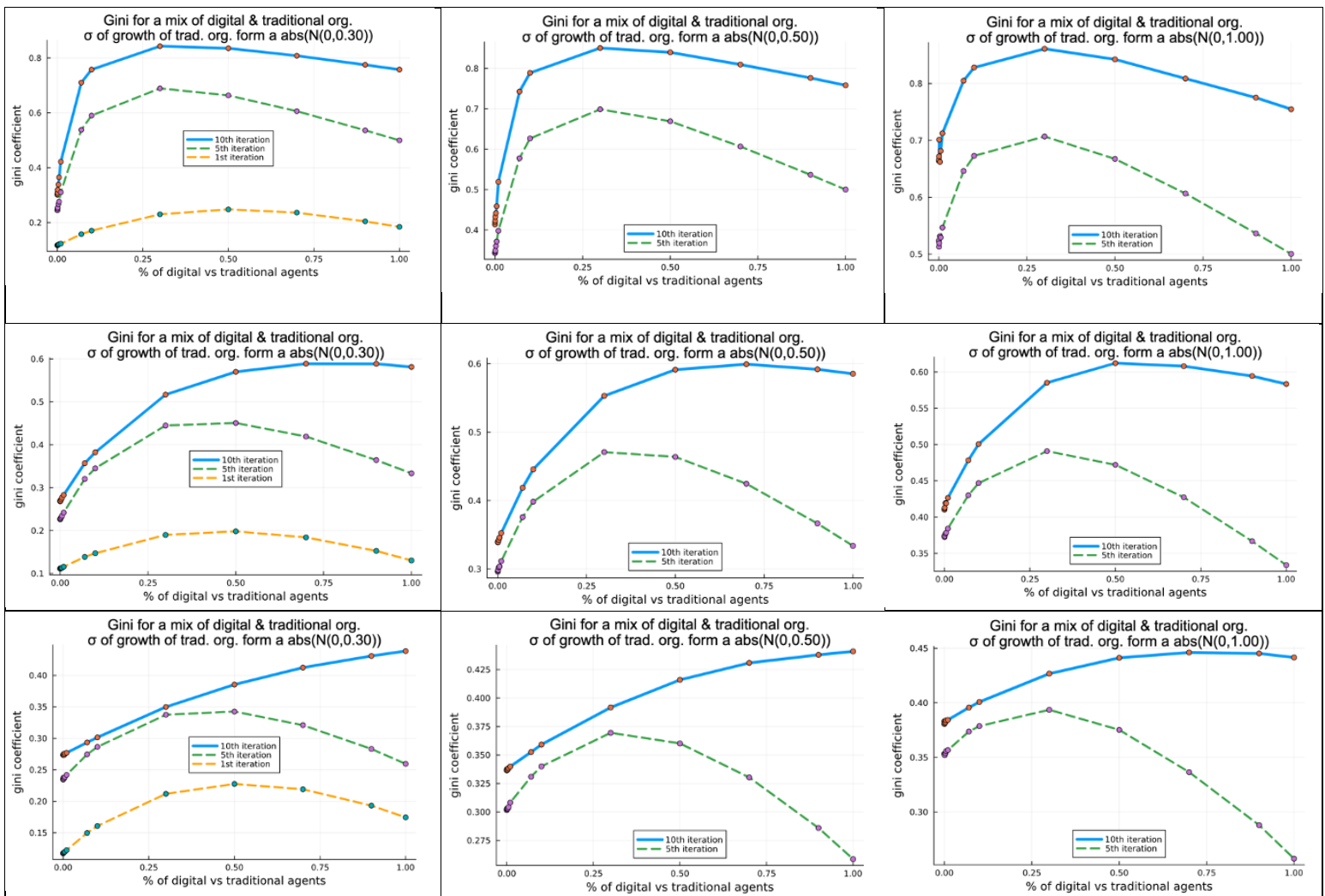

Figure 7. Gini for a mix of traditional and digital companies & different σ for taxations of 10%,15%,25%, 30%; 15%, 25%, 50%, 70% and 10%, 15%, 30% and 90%

Progressive taxation serves as an effective tool for curbing inequality in societies primarily composed of traditional companies, where the magnifying effect of a selection process that values moderate individual contributions is itself progressive and moderate. However, when two distinct types of growth are introduced—particularly when one type is marked by the high individual growth of certain agents—only truly elevated levels of taxation make a noticeable difference. As we can observe only when taxation levels in the last tax bracket reach 70% or 90%, we can appreciate a significant impact. These rates would be hard to implement for any state, but certainly difficult to imagine globally.

This observation aligns with the historical context in which progressive taxation emerged. Originally developed in a world where the distribution of companies leaned towards a normal distribution due to diminishing returns on scale, the progressive taxation system was well-suited to that environment. However, the advent of digital companies has inaugurated a new chapter in growth dynamics, challenging the efficacy of progressive taxation.

Consequently, our findings do not negate our second hypothesis. However, they do underscore the need for policy measures aimed at combating inequality stemming from digital transformations to consider the potential of widespread digital technology adoption. By doing so, it is possible to attenuate the diversity in growth trajectories and, at least partially, address the fundamental causes of pronounced disparities.

# Tech Disruption, a Process of Social Adoption where Inequality is an emergent property

Up to this point, we concentrated on understanding the consequences of market dynamics amplifying inequality and the use of progressive taxation as a moderator. However, technological adoption emerges from a complex social process demanding societal validation. Organizations are willing to undertake the risk of technology adoption based largely on observed advantages among a significant number of peers and competitors. This societal mechanism for adopting technology is increasingly influenced by enhanced connectivity, a byproduct of globalization. We simulate this phenomenon using a very small cohort of digital agents that act as catalysts for technological adoption within increasingly interconnected networks.

The foundational structure of our model remains largely consistent. It incorporates a variety of agents categorized into two distinct groups: the conventional agents whose growth per cycle is derived from a normal distribution (mean 0.15, σ 0.1), and the digital organizations with growth rates also stemming from a normal distribution, albeit with a mean of 1.5 and σ of 1. Notably, in our initial setup, the digital organizations account for a mere 1% of the total—a foundational seed.

These agents operate within a Newman-Watts-Strogratz network continuum, transitioning from a regular to a small-world setup and ultimately to a random configuration, modulated by the parameter ß. One primary objective is to determine the starkly contrasting behaviors the model exhibits based on the underlying network topology, with inequality manifesting as an emergent property of the model.Yet, it is crucial to understand that technology adoption is not a simplistic, linear process. It is inherently social, steered by societal confirmation. In every iteration, agents assess their network peers with higher valuations. If a significant number of them, defined by a threshold, have integrated digital technologies, the agent in question follows suit; otherwise, they remain unchanged.

Moreover, an evolutionary dynamic is at play. Agents that deplete their capital are phased out and replaced by new entrants.

## Agent Behavior

Again, agents are initialized with a minimal valuation, but this time they are situated in a node of a Newman-Watts-Strogratz network of β ranging from 0.001 to 1, therefore covering the space from regular to small-world to random topologies.

Only a seed of digital agents which are situated randomly in the network, concretely 0.01% of the agents.

During each iteration of the simulation:

1. Agents are activated in a stochastic manner, and their valuation either appreciates or depreciates based on their designated type.

2. As in the previous case agents obtain gains or loses following the same rules and pay the corresponding taxes if gains are positive as before.

3. Agents assess their rank in the neighborhood. If they are below the average valuation of their neighborhood, they consider a process of digital transformation.

4. If they are considering a process of digital transformation and the number of digital agents in the neighborhood is higher or equal than a threshold, then they become digital conserving their actual age and therefore subjected to the depreciation corresponding to their age.

Again, if the valuation of any agent is depleted, it is considered defunct and a new agent is created according to the same rules than in the initialization. Therefore, the number of agents remains constant.

Once more, agents commence with a minimal valuation. This round, however, positions them within nodes of a Newman-Watts-Strogatz network, with a β value spanning from 0.001 to 1. This spectrum encompasses topologies from regular to small-world, extending to random configurations. A mere seed, constituting 0.01% of the agents and designated as digital, is randomly dispersed across the network.

During each simulation iteration the following steps are followed:

1. **Activation**: Agents are stochastically activated, with their valuations either increasing or shrinking based on their assigned type.

2. **Financial Fluctuations**: Analogous to the previous model, agents witness gains or losses adhering to established rules. Positive gains incur tax liabilities, as previously delineated.

3. **Neighborhood Assessment**: Agents gauge their financial standing within their proximate network of distance one. Should their valuation lag behind their neighbors' average, they contemplate embarking on a digital transformation journey.

4. **Transformation Decision**: If an agent is mulling over a digital overhaul and the count of neighboring digital agents meets or exceeds a set threshold, the agent transitions to a digital one. While their digital status is revamped, their age remains unchanged, and thus, the accompanying age-related depreciation applies.

As with the prior model, should an agent's valuation run dry, it is rendered "defunct." A replacement agent is subsequently birthed, adhering to the original initialization parameters, ensuring a consistent agent count throughout.

## Computational experiments

In this simulation scenario, our model involves a pool of 1,000 agents. The results we present are derived from the average outcomes of 1,000 individual experiments. Given the diffusion process in play, we extended the number of periods to 100, making appropriate adjustments to the depreciation rate. Each agent is connected to 100 neighbors within a Newman-Watts-Strogatz network, modulated by varying values of β—a parameter pivotal in shaping the network topology.

The entire simulation framework has been programmed using Julia 1.9. This foundation is further bolstered by the integration of the agents.jl library (Datseris, Vahdati, and DuBois, 2021) in conjunction with the graphs.jl library to facilitate network functionalities.

For readers seeking an in-depth understanding of the simulation parameters, the subsequent table provides a comprehensive overview.

| PARAMETER | VALUE |
|---|---|
| Number of agents | 1,000 |
| Number of experiments | 1,000 |
| Number of Periods | 100 |
| Taxes | 15% flat<br>10% 15% 25% 30% per quartile<br>15% 25% 50% 70% per quartile |
| Profits/loses per period | a random number from a normal distribution of $\mu = 0.15$ and $\sigma = 0.1$ for traditional companies and $\mu = 1.5$ and $\sigma = 1$ for digital companies |
| Depreciation for digital companies | 30 periods at a rate of 1/30 per period |
| Base growth rate of digital companies after depreciation | The difference between digital companies' growth and two times the one of traditional companies |
| Initial valuation | Ten units |
| Initial percentage of digital companies | 0..01 (100 digital organizations) |
| Evolutionary mechanism | Agents iterate randomly for the established number of periods. At each period they age 1 and in the case of digital agents depreciate their growth rate. They earn gains/losses according to their different growth rates. If they exhaust their capital, they die and are replaced by a new agent |
| Network where agents are situated | Newman – Watts – Strogatz with ß={0.001, 0.01, 0.05, 0.1, 0.15, 0.2, 0.3, 0.4, 0.5, 0.7, 1} |
| Number of neighbors | 100 |
| Mechanism of adoption | Agents examine all neighbors at distance 1 with a higher valuation than themselves, if |
| | among them digital are equal or higher than the aspiration level threshold, they adopt digital technologies |
| Aspiration level threshold | the closes integer of a random number from a normal distribution with μ= 5 and σ = 0.1 also μ ={2, 7} and σ = {0. 0.5, 1, 1.5, 2, 2.5} are considered |

*Table 3. Parameters of the dynamic model*

## Results

The extent and nature of connectivity play a decisive role in determining patterns of technology adoption and, subsequently, emergent inequality as we can observe in the results of our experiments presented in Fig. 8. Specifically, the connectivity within a network exerts influence over how technologies permeate across agents, impacting the adoption rate and consequent inequalities. Results correspond to a threshold of a normal distribution with a mean of five agents and a standard deviation of 0.1. It's important to note that the stability of these findings is observed after 30-50 iterations, even though the results presented here are based on 100 iterations.

Technology adoption is subdued when connectivity remains limited and agents predominantly cluster within their inherent communities. The resulting economic landscape is one where inequality is tempered, mainly due to the restricted diffusion of technological advantages within insular clusters. Essentially, these close-knit communities exhibit homogeneity in growth rates, thereby keeping disparity in check.

However, as connectivity evolves, transitioning from limited connections to the more intertwined structure of small worlds, there is an observable uptick in the rate of digital adoption. As agents become more interlinked, information and trends disseminate more rapidly. This increasing interconnectivity facilitates the spread of digital technology and, in tandem, amplifies inequality. The expanded connections lead to a more diverse distribution of growth rates, with agents that adopt digital technologies pulling ahead of their peers.

The real tipping point arrives when networks veer towards randomness, specifically when the parameter ß surpasses 0.2. At this juncture, the adoption rate not only increases but engulfs the entirety of the network, causing a widespread shift. The technology percolates across all agents, irrespective of their initial stance or community. Such random networks break down the barriers of insular clusters, making it conducive for sweeping technological adoption and, consequently, pronounced inequality.

In essence, the structure and level of connectivity – average path length - within a network serve as determinants for technology adoption, with cascading effects on emergent economic inequalities. Therefore, we support our third hypothesis.

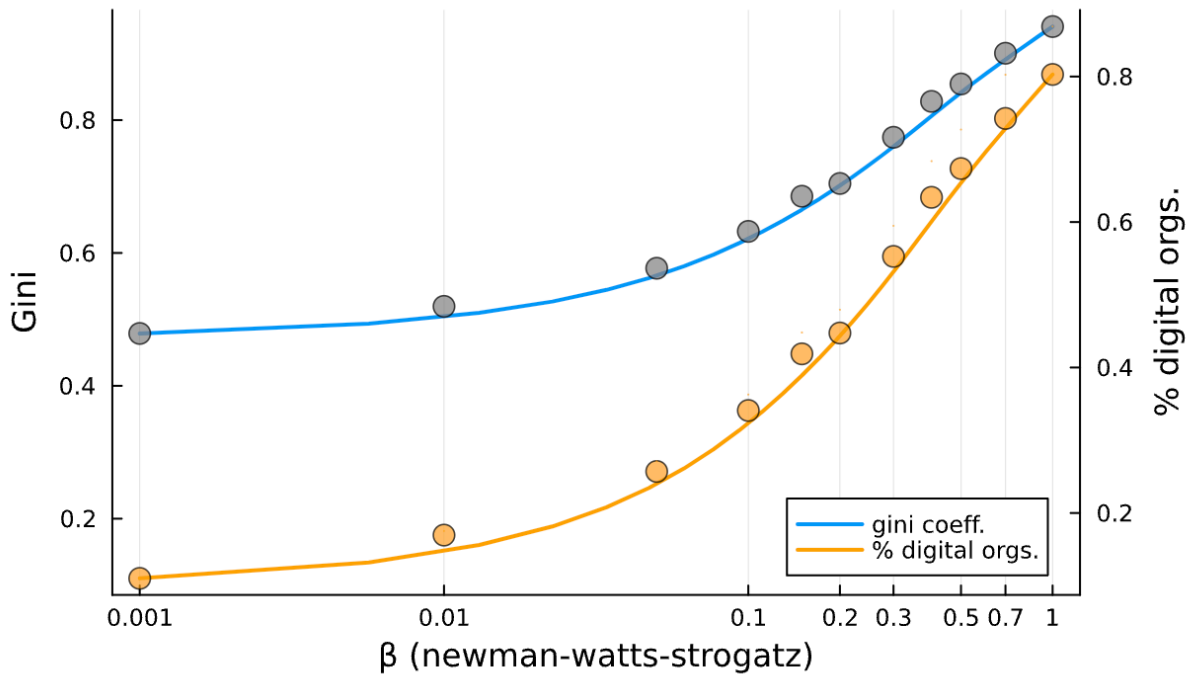

*Figure 8. Dynamic model*

**Impact of progressive taxation**

Once more, our focus shifts to assessing the repercussions of progressive taxation within this model. We adopt an identical scale to the one utilized in the preceding model, as detailed in Table 2.

Initial observations indicate marginal differences between a flat tax rate of 15% and a progressive scheme, starting at 10% in the first quartile and peaking at 30% in the last. These variations manifest minimally in terms of inequality and the proliferation of digital technologies. A discernible deviation emerges only when network topology exhibits clustering with considerable path lengths. Nevertheless, inequality levels harmonize as network connectivity intensifies and average path lengths diminish.

A more pronounced distinction becomes apparent with an escalated tax regimen: an initiation of 15% in the first quartile, culminating at an aggressive 70% in the final quartile. Under this scheme:

1. The uptake of digital technologies remains impervious to the heightened tax slabs, mirroring previous adoption patterns.

2. The Gini coefficient, indicative of inequality, evidences a pronounced decline, especially when β is below 0.2.

The stark reduction in inequality, concurrent with widespread digital tech adoption, defies intuitive expectations. While critics might opine that a top-tier tax rate of 70% could be politically unfeasible in democratic contexts, the underlying implication is significant: it appears feasible to curtail inequality without stifling the adoption of ground breaking technologies. These results shows that taxation certainly has an impact, but limited, adding some nuances to our second hypothesis, however in the same direction.

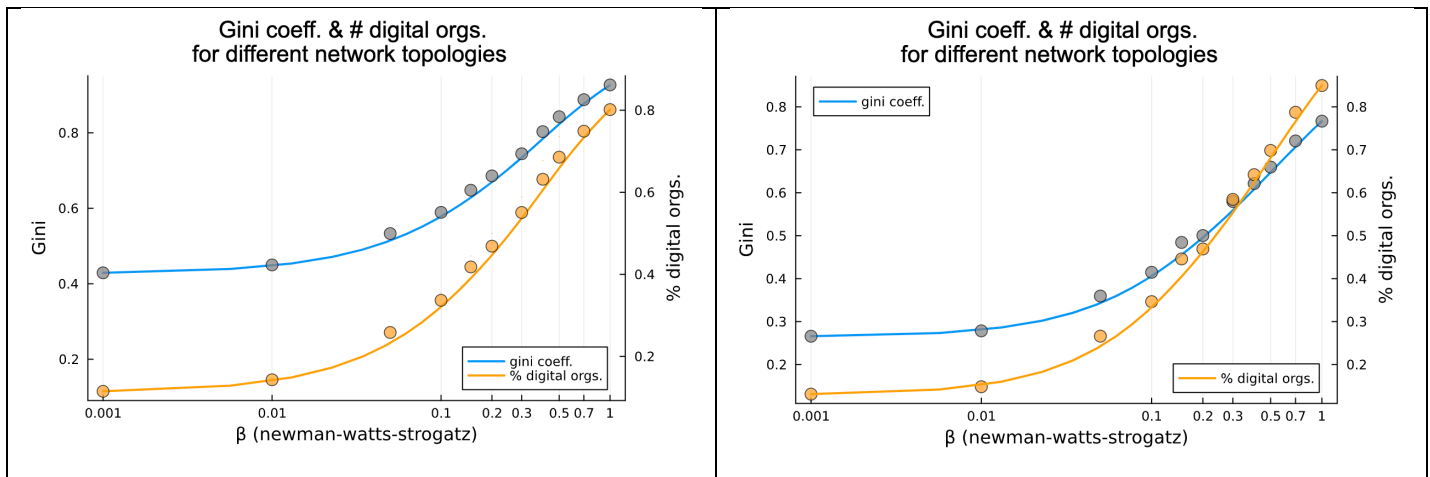

Figure 9.. Impact of progressive taxation on the dynamic model

### How a Few Misfits can Change the World

Drucker's well-noted assertion that "Culture eats strategy for breakfast" underscores the significance of cultural underpinnings in sociotechnical processes, including technology adoption. In our simulations, cultural dynamics are parameterized by the aspiration level threshold. So far, our models employ a broadly consistent threshold, displaying minor deviations among the agent population, specifically with a standard deviation of 0.1. This exemplifies a homogeneous cultural milieu with minimal diversity.

While such homogeneity might have been prevalent in earlier epochs, contemporary societies of the 21st century exhibit pronounced diversity. Nevertheless, certain societies manifest heightened entrepreneurial tendencies – not necessarily because the majority embodies these characteristics, but due to a fraction of audacious entrepreneurs who embrace greater risks.

The importance of this in terms of policy cannot be underscored enough. While effecting widespread cultural change may be beyond the scope of any policy intervention, the possibility of introducing a small cohort of entrepreneurs – a few misfits - and connecting them to global technological hubs is achievable. In our simulations, we model these types of interventions by expanding the σ, leading to a broader distribution of aspiration-level thresholds. Within the constraints of the model, these entrepreneurial agents are offset by a corresponding number of more risk-averse agents, accounting for the worst-case scenario, a counter-reaction, and thereby maintaining the overall population average.

Figure 10 delineates the repercussions of this cultural evolution on our agent-based community. Agents transition from a highly homogeneous aspiration level threshold to a more diverse one, specifically shifting from a standard deviation of 0 to 2.5. With minimal standard deviation, technological shock adoption is scant, representing more conservative societies. However, as the distribution flattens, adoption intensifies, resembling patterns seen with lower thresholds. Therefore, we can support our fourth hypothesis.

A pivotal distinction exists between merely reducing the threshold and amplifying the standard deviation. The latter augments both inequality and adoption across all network topologies, leading to a more uniform adoption curve.

The average path length of the network influences the intricate interplay of adoption and ensuing inequality. However, the diversity in adoption thresholds introduces intriguing dynamics.

This counterintuitive result holds profound implications for innovation policy formulation. While engineering a comprehensive societal cultural shift is challenging, our findings suggest an alternative: Introducing small numbers of change agents. Despite the potential conservative backlash, appears sufficient to galvanize widespread technological adoption.

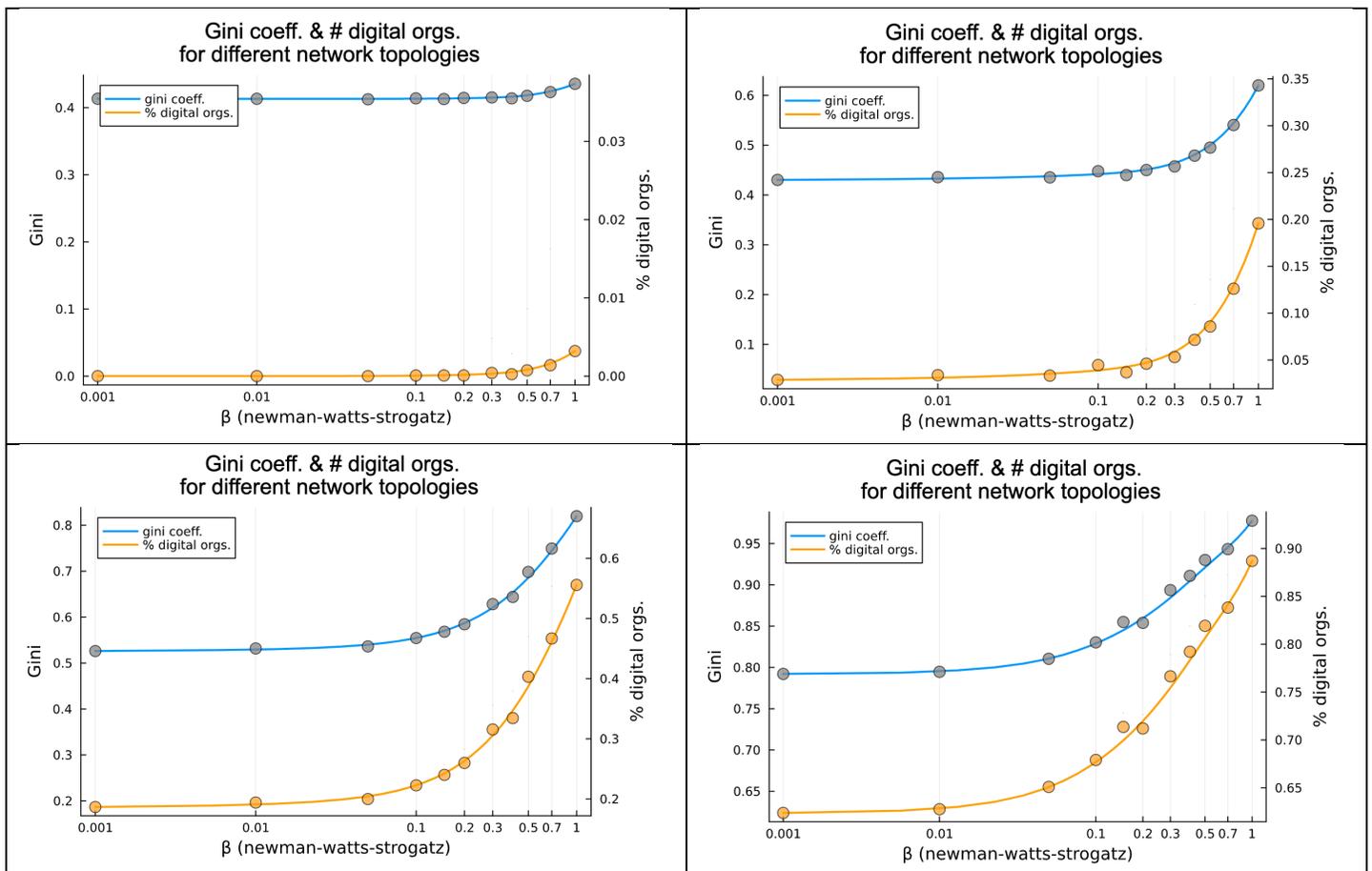

*Figure 10. Gini coeff. & diffusion for aspiration threshold of 9 with σ ={0, 1.5, 2, 2.5}*

## Conclusions

In this paper, we have concentrated on the dual mechanisms through which technological disruptions give rise to inequality. First, we consider the evolutionary dynamics of markets, in which survival of the fittest prevails, often resulting in a polarized population comprising established market leaders and emerging—hence smaller—organizations. Second, we delve into the phenomena of technological diffusion and disruption, conceptualizing them as social adoption processes.

We then examined historical remedies aimed at facilitating wealth redistribution, specifically the role of progressive taxation, assessing its effectiveness in today's hyperconnected global landscape. This environment is dominated by high-growth, high-risk companies characterized by near-zero marginal costs and exceptional scalability. Our analysis reveals the limited impact of progressive taxation, which would need to be significantly escalated as disparities in growth trajectories widen. Such a move is not only challenging to justify ex-ante in democratic societies but also fraught with unintended consequences that could negatively affect the broader economy.

In relation to the influence of market evolutionary dynamics in shaping agent distributions, particularly in contexts of significant growth trajectory disparities, we have elaborated on how these dynamics magnify existing disparities, leading to Pareto-like distributions. Our findings show that just a few iterations are sufficient to significantly alter the market landscape, exacerbating existing inequalities.

These outcomes imply that while progressive taxation may serve as a viable mitigating measure, addressing the root issue—namely the disparities in growth trajectories—is a more effective solution.

Digital disruptions don't emerge ex-ante but evolve through a diffusion process deeply rooted in social construction, where social validation and ensuing bandwagon effects play a pivotal role. Our model reveals that this process is critically influenced by two key factors: system connectivity and an entrepreneurial culture willing to embrace new technologies.

Concerning connectivity, our stable model demonstrates that entering the "small world' zone of the network, characterized by low path lengths, is essential for triggering cascading diffusion. Only then do we attain saturation points that invariably lead to heightened inequality. Our findings indicate that complex contagions, akin to those involved in technology adoption, require network topologies that are conducive to diffusion and adoption. Mere weak ties or simple links are insufficient; wide bridges connecting disparate communities must be established for new technologies to permeate these groups effectively.

While connectivity is a crucial factor, it alone is insufficient for the diffusion of new technologies. A cultural willingness to adopt such technologies is equally indispensable. Changing the prevailing culture, particularly shifting the mean level of risk acceptance, has proven to be a daunting challenge for many societies. However, our research suggests that such sweeping cultural change may not be necessary. Instead, flattening the distribution—by introducing a small cohort of entrepreneurs – a few misfits - willing to take risks, can generate the necessary social validation that facilitates widespread adoption.

This type of policy intervention is well within the grasp of many societies, achievable through acceleration programs, business schools, and initiatives geared towards innovation and entrepreneurship, often manifested in the form of specialized 'factories' or 'programs.' While such interventions may not fully compensate for limited network connectivity, they can significantly accelerate technology adoption within societies, leading to bandwagon effects in technology uptake.

In summary, while loading the economy with high levels of taxation on largely diverse growth trajectories may mitigate inequality, it comes at the substantial cost of dampening overall economic growth and offers limited long-term efficacy. Tackling the root issue—the disparity in these growth trajectories—by accelerating technology adoption provides a more enduring and sustainable impact. Our research shows that targeted interventions can be transformative; introducing just a few risk-takers – a few misfits - into the system can indeed change the world.